\documentclass[fleqn,usenatbib]{mnras} %a4paper

\usepackage{newtxtext,newtxmath}
\usepackage[T1]{fontenc}
\usepackage{ae,aecompl}

\usepackage{graphicx}	% Including figure files
\usepackage{amsmath}	% Advanced maths commands
\usepackage{grffile}    % allows "dots" in filenames (e.g. m1.5_c3.png)
\usepackage{breqn}
\def\Mdot{\hbox{${\dot M}$}}
\def\Edot{\hbox{${\dot E}$}}

\def\cm{{\rm\thinspace cm}}
\def\km{{\rm\thinspace km}}
\def\s{{\rm\thinspace s}}
\def\yr{{\rm\thinspace yr}}
\def\g{{\rm\thinspace g}}

\def\kmps{\hbox{${\rm\km\s^{-1}\,}$}}

\def\gpcm3{\hbox{${\rm\g\cm^{-3}\,}$}}

\def\Msol{\hbox{${\rm\thinspace M_{\odot}}$}}

\def\Msolpyr{\hbox{${\rm\Msol\yr^{-1}\,}$}}
\def\spose#1{\hbox to 0pt{#1\hss}}
\def\ltsimm{\mathrel{\spose{\lower 3pt\hbox{$\sim$}}
        \raise 2.0pt\hbox{$<$}}}
\def\gtsimm{\mathrel{\spose{\lower 3pt\hbox{$\sim$}}
        \raise 2.0pt\hbox{$>$}}}

\title[How to inflate a wind-blown bubble] %max 50 characters
{How to inflate a wind-blown bubble}

\author[J.~M.~Pittard et al. ]{J.~M.~Pittard\thanks{E-mail:
    j.m.pittard@leeds.ac.uk}, C.~J.~Wareing and M.~M.~Kupilas\\
School of Physics and Astronomy, University of
       Leeds, Woodhouse Lane, Leeds LS2 9JT, UK\\   
}

\date{Accepted 2021 September 13. Received 2021 September 13; in original form 2021 July 30}

\pubyear{2021}

\begin{document}
\label{firstpage}
\pagerange{\pageref{firstpage}--\pageref{lastpage}}
\maketitle

% Abstract of the paper (max 250 words)
\begin{abstract}
  Stellar winds are one of several ways that massive stars can affect
  the star formation process on local and galactic scales. In this
  paper we investigate the numerical resolution needed to inflate an
    energy-driven stellar wind bubble in an external medium. We find
  that the radius of the wind injection region, $r_{\rm inj}$, must be
  below a maximum value, $r_{\rm inj,max}$, in order for a bubble to
  be produced, but must be significantly below this value if the
  bubble properties are to closely agree with analytical
    predictions. The final bubble momentum is within 25 per cent of
    the value from a higher resolution reference model if
    $\chi = r_{\rm inj}/r_{\rm inj,max}$ = 0.1. Our work has
  significance for the amount of radial momentum that a wind-blown
  bubble can impart to the ambient medium in simulations, and thus on
  the relative importance of stellar wind feedback.
\end{abstract}


%%%%%%%%%%%%%%%%%%%%%%%%%%%%%%%%%%%%%%%%%%%%%%%%%%

%%%%%%%%%%%%%%%%% BODY OF PAPER %%%%%%%%%%%%%%%%%%

\section{Introduction}
\label{sec:intro}
Massive stars have dramatic impacts on their surroundings, through
their intense radiation, and their powerful winds and supernova
explosions. These stellar inputs rapidly destroy the molecular clouds
in which stars form, and are also able to affect the global structure
and evolution of their host galaxy. The momentum that is injected into
the interstellar medium (ISM), plus the boost through $PdV$ work done
by over-pressured expanding gas, determines the amplitude of the
turbulent gas motions, which limit gravitational condensation and
collapse, and ultimately limit and regulate star formation
\citep[e.g.,][]{Shetty:2012}.

Recent work has indicated that early (pre-supernova) feedback from
winds and radiation is required to explain the anti-correlation of
giant molecular clouds (GMCs) and ionized regions on 100\,pc scales
and less
\citep[e.g.,][]{Kruijssen:2019,Chevance:2020,Chevance:2021}. GMCs
appear to disperse within $1-5$\,Myr of massive stars emerging from
their natal clouds, with photoionization and stellar winds seeming to
play a crucial role. Cosmological simulations with only supernova (SN)
feedback also show that SNe alone are not able to prevent excessive
star formation \citep*[e.g.,][]{Smith:2019}.

Nonetheless, the impact of stellar winds remains much debated.
\citet{el-badry:2019} showed that turbulent mixing at the interface
between the hot interior gas and colder exterior gas sets the cooling
losses, which reduces the radial momentum by a factor of 2. However,
the applicability of the classical energy-conserving wind-blown-bubble
approach \citep{Weaver:1977} has recently been questioned by
\citet{Lancaster:2021a,Lancaster:2021b}, who argue that if the
interface becomes fractal-like due to the presence of inhomogenities
in the ambient gas, radiative losses can become very substantial and
cause the bubble to display momentum-conserving-like
behaviour. \citet{Dinnbier:2020} argued that stellar winds and
photoionization are quenched in clusters with a mass above
$10^{4}\,\Msol$.

Many works simply assume that the feedback from winds is momentum-driven
\citep[e.g.,][]{Dale:2014}, which sets a lower limit to their
impact. On the other hand, if stellar winds couple relatively weakly
to the densest clumps of gas in star-cluster environments, the winds
can carve and open up low-density channels within their environments
\citep[e.g.,][]{Rogers:2013,Wareing:2017b} and thus are still able to create
wind-blown bubbles with low density hot interiors, that might be
capable of doing significant $PdV$ work. Stellar wind feedback may also
be shaped by the large scale density distribution arising from a
large-scale magnetic field \citep[e.g.,][]{Wareing:2017a,Wareing:2018} or gas
motions.

The significant momentum boost that can be provided by a wind-blown
bubble is key to having strong stellar wind feedback. However, to correctly
determine the momentum boost requires that simulations have a certain
numerical resolution that we find has not always been achieved in the
literature. In this work we examine the different ways that a wind can
be initiated and also how the development of the wind-blown bubble
depends on numerical resolution. We focus only on the effect of the
wind, so that other effects due to photoionization, for example, do
not complicate the matter.  In Sec.~\ref{sec:wbbs} we discuss the
essential theory of wind-blown bubbles. In Sec.~\ref{sec:model} we
describe our numerical model and the implentation of the wind driving,
and in Secs.~\ref{sec:results} and~\ref{sec:discuss} we present and
discuss our results. We summarize and conclude in
Sec.~\ref{sec:summary}.

\section{Wind-blown bubbles}
\label{sec:wbbs}
\subsection{Essential features}
The essential features of an idealised spherically-symmetric wind-blown bubble are (moving outwards
from the wind source) an inner region
where the wind is freely expanding, a region of shocked wind, and a region
of swept-up ambient material. The shocked and unshocked wind regions
are separated by a reverse shock (RS), the wind and ambient gas is
separated by a contact discontinuity (CD), and the swept-up material
is bounded by a forward shock (FS). Typically, the shocked wind is at
much lower density than the ambient gas and remains hot with long
cooling timescales, whereas the swept-up gas cools efficiently and is
compressed into a cooler shell. The radius of an adiabatic wind-blown
bubble (where there is no cooling of the shocked wind) with a thin
swept-up shell is \citep[e.g.,][]{Dyson:1980}
\begin{equation}
\label{eq:rFS}  
r_{\rm bub} = \left(\frac{125}{154\pi}\right)^{1/5}
\left(\frac{\Edot}{\rho_{\rm amb}}\right)^{1/5} t^{3/5},
\end{equation}
where $\Edot$ is the rate of energy injection of the wind
($\Edot = \frac{1}{2}\Mdot \,v_{\rm w}^{2}$, where $\Mdot$ is the
mass-loss rate of the star and $v_{\rm w}$ is the terminal speed of
the wind), $\rho_{\rm amb}$ is the density of the ambient medium and
$t$ is the bubble age. This equation is valid when the pressure of the
bubble is much greater than the pressure of the ambient gas \citep[for
solutions when this is not the case see][]{Garcia:1996}. We define the
radius of the reverse shock and contact discontinuity as $r_{\rm rs}$
and $r_{\rm cd}$, respectively.

The bubble is initially of zero size
($r_{\rm rs} = r_{\rm cd} = r_{\rm bub} = 0$ at $t=0$). All 3 radii
then increase with time. The bubble will expand as long as its
interior pressure exceeds the ambient pressure, $P_{\rm amb}$, or as
long as the lifetime of the source. The thermal pressure within the
bubble is \citep[e.g.,][]{Dyson:1980,Pittard:2013}
\begin{equation}
P_{\rm bub} = \frac{7}{(3850\pi)^{2/5}}\Edot^{2/5}\rho_{\rm amb}^{3/5}t^{-4/5}.  
\end{equation}  

At all times the position of the reverse shock is set by pressure balance between
the ram pressure of the hypersonic wind (in the frame of the reverse
shock) and the thermal pressure of the hot bubble:
\begin{equation}
\label{eq:wbbPressBal}  
\rho_{\rm w} (v_{\rm w} - v_{\rm rs})^{2} \approx P_{\rm bub},  
\end{equation}  
where $\rho_{\rm w}$ is the pre-shock density of the wind at the
reverse shock and $v_{\rm rs}$ is the velocity of the reverse shock.
If the radial velocity of the reverse shock is ignored, the
reverse shock position is given by
\begin{equation}
\label{eq:rRS}
r_{\rm rs} \approx \left(\frac{\Mdot v_{\rm w}}{4\pi P_{\rm
      bub}}\right)^{1/2} \approx 0.70\left(\Mdot v_{\rm w}\right)^{1/2}
\dot{E}^{-1/5}\rho_{\rm amb}^{-3/10} t^{2/5}.
\end{equation}
The reverse shock will be slightly closer to the star if
$v_{\rm rs}/v_{\rm w}$ is significant (e.g., at early times). The
reverse shock attains a maximum radius, $r_{\rm rs,max}$, when
$P_{\rm bub}=P_{\rm amb}$. This is given by
\begin{equation}
\label{eq:rRS_max}
r_{\rm rs,max} \approx \left(\frac{\Mdot v_{\rm w}}{4\pi P_{\rm
      amb}}\right)^{1/2},
\end{equation}
and occurs when the bubble age is
\begin{equation}
\label{eq:rRS_max_t}
t_{\rm rs,max} = 0.104\,\Edot^{1/2}\,\rho_{\rm amb}^{3/4}\,P_{\rm amb}^{-5/4}.
\end{equation}
Given that $t_{\rm rs,max}$ may be (much) greater than the lifetime of
the star, a more useful measure than $r_{\rm rs,max}$ is the
radius of the reverse shock at the end of life of the star, which we
define to be
\begin{equation}
\label{eq:rRS_tlife}
r_{\rm rs,tlife} \approx 0.70\left(\Mdot v_{\rm w}\right)^{1/2}
\dot{E}^{-1/5}\rho_{\rm amb}^{-3/10} t_{\rm life}^{2/5},
\end{equation}
where $t_{\rm life}$ is the lifetime of the star. In fact, unless the
wind is very weak and the ambient pressure is very high, we always
expect $t_{\rm life} < t_{\rm rs,max}$ (see
  Sec.~\ref{sec:results} for typical values of $t_{\rm rs,max}$ and $t_{\rm life}$).

\subsection{Momentum injection}
The wind injects momentum at a rate
$\dot{p}_{\rm wind}=\Mdot v_{\rm w}$, so the momentum supplied by the
wind,
\begin{equation}
\label{eq:mtmwind}
p_{\rm wind}=\dot{p}_{\rm wind}\,t.
\end{equation}
However, the $PdV$ work by the
bubble on the surrounding gas means that the momentum of the bubble
(which is dominated by the swept-up shell) is
\begin{equation}
\label{eq:mtmbub}
p_{\rm bub} = \frac{4\pi}{3}r_{\rm bub}^{3}\rho_{\rm amb}\dot{r}_{\rm
  bub} = 0.85\, \dot{E}^{4/5} \,\rho_{\rm amb}^{1/5} \,t^{7/5}.
\end{equation}
The momentum boost provided by the bubble is
\begin{equation}
\label{eq:mtmboost}
\beta = \frac{p_{\rm bub}}{p_{\rm
  wind}} = 0.60\, \dot{E}^{3/10} \,\dot{M}^{-1/2} \,\rho_{\rm amb}^{1/5} \,t^{2/5}.
\end{equation}
$\beta$ can easily have a value in excess of 100.

Since the wind momentum increases linearly with time, while the
wind-blown bubble momentum increases as $t^{7/5}$, at early times the
wind will have more momentum than the bubble. This non-sensical result
indicates a break-down of the bubble model at early times. It arises
because the bubble has not existed long enough to have properly
developed its characteristic features. We define the time at which $p_{\rm
  wind}=p_{\rm bub}$ as 
\begin{equation}
\label{eq:wbb_teq}
t_{\rm eq} = 3.57 \,\Mdot^{5/4} \,\Edot^{-3/4} \,\rho_{\rm amb}^{-1/2}.  
\end{equation}

\section{The numerical model}
\label{sec:model}

There are a number of different ways that a hypersonic stellar wind
can be modelled using a grid-based hydrodynamics code. In all cases
several cell-averaged quantities inside of a ``remap'' or
``injection'' radius, $r_{\rm inj}$, are altered or reset at each
timestep, $dt$. The volume of the injection region is $V_{\rm
  inj}$. Typically $r_{\rm inj}$ and $V_{\rm inj}$ are fixed, but
there is no reason why they might not instead be time dependent (e.g.,
if the grid adaptively derefines or is expanding - as we show
later).

In the following we shall assume for simplicity that the wind
is spherically symmetric and that it blows into a static medium with a
uniform density and pressure (though this is often not the case in
reality). We also assume that there is no non-thermal contribution to the
  ambient pressure. Table~\ref{tab:models} notes details of our models and the final momentum and
momentum boost that is obtained.

\subsection{Some possible implementations}
\label{sec:wind_launch_methods}
\subsubsection{Momentum and energy overwrite (method {\em meo})}
\label{sec:meo}
In this scenario, which is perhaps the simplest approach, the density
and velocity of each cell with $r < r_{\rm inj}$ are set to values
appropriate for a free-flowing wind. In the case of a spherically
symmetric wind, if the inner and outer radius of
the cell are $r_{\rm i}$ and $r_{\rm o}$, respectively, then the mass
within the cell is
\begin{equation}
M = \int_{r_{\rm i}}^{r_{\rm o}} 4\pi r^{2}\rho \,dr =
\frac{\Mdot}{v_{\rm w}} \left(r_{\rm o} - r_{\rm i}\right).
\end{equation}  
Since the volume of the cell is $\frac{4}{3}\pi (r_{\rm o}^{3}-r_{\rm
  i}^{3}$), the cell density
\begin{equation}
\rho = \frac{3\Mdot}{4\pi v_{\rm w}}\frac{\left(r_{\rm o} - r_{\rm i}\right)}{\left(r_{\rm o}^{3}-r_{\rm
  i}^{3}\right)}.  
\end{equation}  
The cell velocity ${\bf v}=v_{\rm w}$. The existing values in each cell are overwritten at each
time step and previous information about the flow within the injection
region is lost. The pressure in the cell is set so that the cell
temperature is at a desired value (e.g., $10^{4}\,$K, or the floor
temperature of the simulation, $T_{\rm floor}$). As long as the sound
speed in the wind is much less than the wind speed (i.e. the wind is
hypersonic), the exact temperature of the wind will not be important
as the kinetic energy dominates. In this procedure the wind energy is
almost purely kinetic, and the density within the injection region
falls as $1/r^{2}$.

\subsubsection{Thermal energy injection (method {\em ei})}
\label{sec:ei}
Another possibility is to inject the wind energy as purely thermal. The mass and energy
injection from the wind are shared uniformally for cells with $r <
r_{\rm inj}$. The procedure, which is applied to each cell within
the injection region at every time step is \citep{Chevalier:1985,Wunsch:2008}:

\begin{enumerate}
\item Add mass to the cell: $\rho_{\rm new} = \rho_{\rm old} + d\rho$,
  where $d\rho = \Mdot dt/V_{\rm inj}$.
\item Conserve momentum: ${\bf v}_{\rm new} = {\bf v}_{\rm old} \rho_{\rm
    old}/\rho_{\rm new}$.
\item Calculate the new kinetic energy density in the cell and subtract from
  the old total energy density, $e_{\rm tot,old}$, to
  give the new internal energy density prior to the addition of the new
  (thermal) energy:
    $e_{\rm int} = e_{\rm tot,old} - \rho_{\rm new}{\bf v}_{\rm
      new}^{2}/2.$
\item Add the new (thermal) energy to the new internal energy density: $e_{\rm int,new} =
  e_{\rm int} + \Edot dt/V_{\rm inj}$.
\end{enumerate}
In the above prescription, the old and new values in the cell have the
subscript ``old'' and ``new'', respectively. 

With this procedure the flow is thermally driven, and transitions from
subsonic to supersonic at the edge of the injection region. Outside of
the injection region the flow continues to accelerate due to the
thermal pressure gradient, and asymptotically reaches its terminal
speed of $v_{\rm w} = \sqrt{2\Edot/\Mdot}$ at large radii. Because
cell quantites within the injection region retain some element
of their previous values, the resultant flow has some sensitivity to
the initital parameters and may develop differently in certain
situations.  For instance, if the initial gas density within the
injection region is very high, rapid cooling of the gas may suppress
development of the wind. If cooling is not significant, the
temperature of the injection region gradually increases until a
stationary flow is produced.

This prescription has the drawback that at early times velocities and
temperatures within the injection region may be low. Some additional
constraints on $dt$ other than the dynamics may then be necessary
(e.g., a cooling time constraint).

\subsubsection{Momentum and energy injection (method {\em mei})}
\label{sec:mei}
Another possibility is to inject mass, momentum and energy evenly into all
cells \citep{Geen:2021}. The procedure is:
\begin{enumerate}
  \item Add mass to each cell: $\rho_{\rm new} = \rho_{\rm old} +
    d\rho$.
  \item Add momentum to each cell: ${\bf v}_{\rm new} = (\rho_{\rm
      old}{\bf v}_{\rm old} + d\rho{\bf v}_{\rm w})/\rho_{\rm new}$.
\item Add energy to each cell: $e_{\rm tot,new} = e_{\rm
    int,old}+0.5\rho_{\rm old}{\bf v}_{\rm old}^{2}+de$, where $de = 0.5 d\rho
 {\bf v}_{\rm w}^{2}$, the old internal energy
 density is $e_{\rm int,old}$, and $e_{\rm tot,new}$ is the new total energy density. 
\item Although not explicitly stated by \citet{Geen:2021}, in order to conserve energy the internal energy density of the cell must become
  $e_{\rm int,new} = e_{\rm tot,new} - 0.5 \rho_{\rm new}{\bf v}_{\rm new}^{2}$.
\end{enumerate}

In this scenario, the stationary flow develops so that the density and
pressure within the injection region increase with radius (in contrast
to method {\em ei} where these quantities decline), and the velocity
of gas within the injection region is everywhere equal to the wind
speed (like the fixed speed overwrite procedure in
Sec.~\ref{sec:meo}). Like the method in Sec.~\ref{sec:ei}, this method
has some sensitivity to the initial conditions, but has the potential
advantage that it may force shorter dynamical timesteps early in the
simulation.

\subsection{Resolution requirements}
\label{sec:res_requirements}
In order for the wind to have {\em any} chance of inflating a bubble
using the {\it meo} wind injection method, the wind ram pressure
at the edge of the injection region must exceed the ambient
pressure. Alternatively, if the {\it ei} wind injection method is
used, the central pressure in the injection region should exceed
$P_{\rm amb}$. Both requirements result in essentially the same
maximum size of the injection region,
\begin{equation}
\label{eq:rinj_max}  
r_{\rm inj,max} = \left(\frac{\Mdot \,v_{\rm w}}{4 \pi P_{\rm amb}}\right)^{1/2},
\end{equation}
above which we do not expect a bubble to be created. We define
\begin{equation}
\label{eq:chi}  
\chi = \frac{r_{\rm inj}}{r_{\rm inj,max}}.
\end{equation}  
Simulations with $\chi < 1$ should inflate a bubble. However, if this
is only marginally satisfied, we do not expect the resulting bubble to
match the analytical solution. In such a scenario the bubble would not experience
such high initial pressures as seen in better resolved bubbles early in their
life - as a result they will evolve to be too small
with too little radial momentum.

In order not to miss {\em any} initial momentum boost with the {\it
  meo} injection method, the initial conditions should be set so that
the momentum (in the injection region) of the outflowing wind (which
has a flow-time or age $t=r_{\rm inj}/v_{\rm w}$) is substantially
greater than the momentum of a wind-blown bubble of equivalent age
(i.e. $t \ll t_{\rm eq}$ - see Eq.~\ref{eq:wbb_teq}). This sets the
constraint $r_{\rm inj} \ll v_{\rm w} t_{\rm eq}$. In essence, this
requirement ensures that the simulation starts prior to the bubble
generating additional momentum through $PdV$ work.

\subsection{The calculations}
\label{sec:calcs}
We are investigating the evolution of a wind-blown
bubble. It makes sense, therefore, to use a code where the grid can
expand with time (if desired). We therefore make use of a heavily
modified version of
VH-1\footnote{http://wonka.physics.ncsu.edu/pub/VH-1/}. The standard inviscid
equations of hydrodynamics in conservative lagrangian form are solved
on a spherically symmetric one-dimensional grid. Piecewise parabolic spatial
reconstruction is used to calculate the interface values. The updated
quantities are then remapped to the original (or an expanding) grid at the end of each
step (this is the piecewise parabolic method (PPM) with lagrangian
remap (``PPMLR'') approach used by VH-1). We use a courant number of 0.6.

Gas can heat and cool via
operator splitting. The rate of change of the internal energy
per unit volume is given by
\begin{equation}
 \dot{e} = n\Gamma - n^{2}\Lambda, 
\end{equation}  
where $n=\rho/m_{\rm H}$ and $\Gamma$ and $\Lambda$ are heating and
cooling coefficients. In this work we assume that
$\Gamma = 2 \times 10^{-26} \,{\rm erg\,s^{-1}}$
(independent of $\rho$ or $T$). The
cooling coefficient, $\Lambda(T)$, is detailed in
\citet{Wareing:2016}. The low temperature part ($T \leq 10^{4}\,$K) is
a corrected fit to the data in \citet{Koyama:2000}. Between
$10^{4}-10^{7.6}$\,K cooling rates calculated with CLOUDY v10.00
\citep{Gnat:2012} are used, while cooling rates calculated from the
MEKAL plasma emission code \citep[][as distributed in XSPEC
v11.2.0]{Kaastra:1992,Mewe:1995} are used for $T\geq10^{7.6}$K.

We restrict cooling at unresolved interfaces between hot diffuse gas
and cold dense gas by replacing the change in the internal energy
density, $de$, with the minimum of the neighbouring $de$'s
at the interface. The cooling is sub-cycled so that the cooling curve
is always sampled with a temperature resolution of at least 20 per
cent.

We also assume solar abundances for the gas \citep[mass fractions
$X_{\rm H}=0.7381$, $X_{\rm He}=0.2485$, and
$X_{\rm Z}=0.0134$;][]{Grevesse:2010}, and a temperature-dependent
average particle mass, $\mu$, is used. In the molecular phase
$\mu = 2.36$, reducing to 0.61 in ionized gas. The value of $\mu$ is
determined from a look-up table of values of $p/\rho$
\citep{Sutherland:2010}. A temperature-independent value of $\gamma$,
the ratio of specific heats, is used, which we set to $\gamma = 5/3$
(see \citet*{Krumholz:2007} for arguments as to why this is also
appropriate for low temperature molecular gas).

\section{Results}
\label{sec:results}
In all of our calculations the stellar wind and ambient medium parameters
are set to $\Mdot = 10^{-7} \Msolpyr$, $v_{\rm w}=2000\, \kmps$,
$\rho_{\rm amb}=2\times10^{-21} \,{\rm g\,cm^{-3}}$ and
$P_{\rm amb} = 1.48\times10^{-12} \,{\rm dyn\,cm^{-2}}$
($P_{\rm amb}/k = 1.07\times10^{4}\,{\rm K\,cm^{-3}}$). These are
typical of a massive hot star on the main sequence and the cold
molecular medium. However, we expect our results to be
applicable to a wide range of wind and ISM parameters, including for
example superbubbles. The ambient medium has an equilibrium temperature
$T_{\rm amb}=21$\,K and average particle mass
$\mu_{\rm amb} = 2.36$. The ionized wind material has an average
particle mass $\mu_{\rm w} = 0.61$.

Our chosen parameters give $r_{\rm rs,max}=2.8\,$pc
(Eq.~\ref{eq:rRS_max}) at $t_{\rm rs,max}=215$\,Myr. However, we only
evolve the simulations for 5\,Myr, this being more typical of the
lifetime of a massive star with the adopted wind parameters.  At this
time we expect the reverse shock to have a radius
$r_{\rm rs,tlife}\approx 0.6$\,pc (see Eq.~\ref{eq:rRS_tlife}), and
for the bubble to still be overpressured with respect to the ambient
medium by a factor of $\approx 20$. Our parameters also give
$r_{\rm inj,max} = 2.68$\,pc.

We use 10 grid cells for the injection region in all of our
calculations. We do not expect our results to be very sensitive to the
exact number of cells in this region, but a mimimum of 4 is probably a
good idea.

\subsection{A model with ``high'' resolution} 
We begin by using the method in Sec.~\ref{sec:meo} to setup and
continue blowing the wind; i.e. the momentum and energy within the 10
closest cells to the grid origin is overwritten after every step. 

For our chosen parameters, $t_{\rm eq} = 1.2\times10^{8}$\,s
($3.8$\,yr). In this time a freely expanding wind will have blown out
to a radius $r_{\rm eq} = v_{\rm w}\,t_{\rm eq} = 0.008\,$pc. To
satisfy the constraint that $r_{\rm inj} \ll r_{\rm eq}$, we use a
grid with 1200 cells, with an initial uniform cell width
$dr_{0}=10^{-5}\,$pc. The wind initially extends out to the edge of
the injection region ($r_{\rm inj}=10^{-4}$\,pc). The flow time of the
wind to this radius is $t = 1.54\times10^{6}\s$, which is
$\ll t_{\rm eq}$, as required.

The average wind density in the last cell in the injection region is
initially $2.9\times10^{-20}\gpcm3$, which is more than 10 times the
density of the ambient gas. The impact of the wind on the ambient gas
creates two shocks, which in this case both move outwards on the grid
(with a density below $\approx 2.2\times10^{-23}\gpcm3$ in the final
cell in the injection region the reverse shock initially moves towards
the grid origin and into the injection region - it is vital that this
is avoided otherwise some mass, momentum and energy is lost from the
simulation).

We fix the size of the grid for a time
$t_{\rm fix}=(r_{\rm max,0}-r_{\rm inj})/v_{\rm w}$, where the maximum
grid radius is initially $r_{\rm max,0}=1.2\times10^{-2}\,$pc.  After
this time we make the grid expand at a rate such that
$r_{\rm max} = r_{\rm max,0}\, (t/t_{\rm fix})^{3/5}$. As the grid
expands, so does the injection region. This has no effect on the
results provided that the reverse shock remains outside of the
injection region, and we can confirm that this constraint is satisfied
at all times.  We set the time at the start of the simulation to
$t=0$. After 5\,Myr of evolution, the grid has expanded to the extent
that $dr = 0.36$\,pc and $r_{\rm max} = 43.6$\,pc. We refer to this
model as ``{\it modx}'' (``x'' for ``expansion'') for the
remainder of this paper, and we adopt it as our reference simulation.

\begin{figure}
\includegraphics[width=8.0cm]{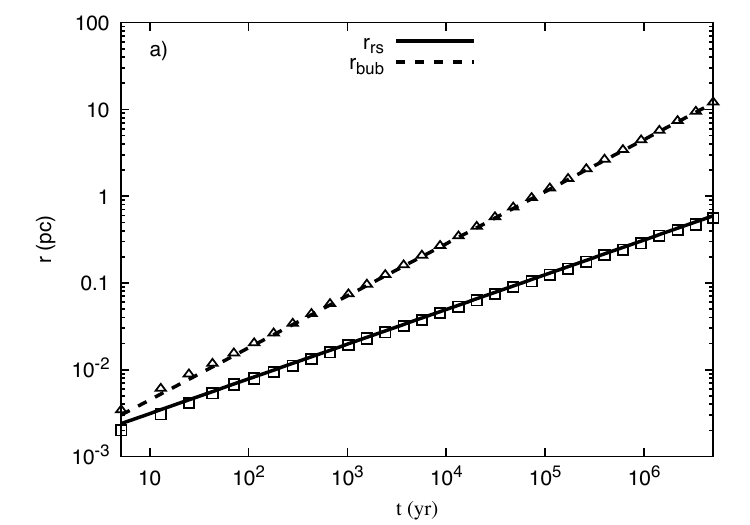}
\includegraphics[width=8.0cm]{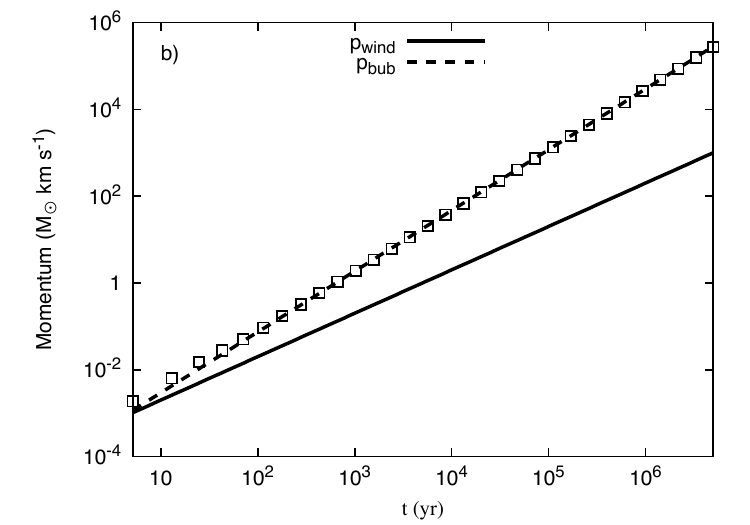}
\caption{a) The radii of the reverse shock and forward shock, and b)
  the momentum of the bubble, as a function of bubble age. Data from
  model {\it modx} is shown by the points, while the lines show the
  analytical values for the shock positions (Eqs.~\ref{eq:rRS}
  and~\ref{eq:rFS}) and for the integrated wind and bubble momenta
  (Eqs.~\ref{eq:mtmwind} and~\ref{eq:mtmbub}). }
\label{fig:wbb1}
\end{figure}

The radii of the forward and reverse shocks as a function of time are
shown in Fig.~\ref{fig:wbb1}a). Both shocks move steadily outwards. They
are reasonably close together when the bubble is young, but since the
bubble expands faster than the reverse shock, most of the bubble
becomes occupied by hot shocked stellar wind material at later
times. The radii measured from the numerical simulation agree very
well with the analytical expectations ($r_{\rm rs}$ and $r_{\rm bub}$).

The momentum of the bubble and the integrated injected momentum of the
wind are shown in Fig.~\ref{fig:wbb1}b). At very early times the
momentum measured from the simulation is slightly above the expected
value from the \citet{Weaver:1977} model, but within 100\,yrs the two
obtain excellent agreement and then continue to do so. The bubble
momentum reaches a maximum of $2.76\times10^{5}\,\Msol \kmps$ at
$t=5$\,Myr, giving a momentum boost $\beta = 276$. The final momentum
is within 2\% of the value predicted by the \citet{Weaver:1977} model
($2.81\times10^{5}\,\Msol \kmps$). This difference is due to the small
amount of cooling that takes place in the shocked stellar wind in the simulation.

\subsection{The effect of numerical resolution}
We now investigate how the bubble momentum changes if we change the
numerical resolution of our simulation. We know that we will miss some
of the momentum boost that the bubble provides if we do not satisfy
$t \ll t_{\rm eq}$, and we want to explore how this loss varies with
$\chi$. To do so we run simulations with a fixed (non-expanding)
grid. We begin with a resolution $dr = 0.025$\,pc. With the standard
10 cell injection region the edge of the injection region is at
$r_{\rm inj}=0.25$\,pc, giving $\chi=0.093$. This resolution should be
just about high enough for the simulated bubble to match the
analytical predictions reasonably well. We refer to this model as
``{\it meo_0.1}'' (``{\it meo}'' for the wind launch mechanism
(momentum and energy overwrite) and ``{\it 0.1}'' for the value of
$\chi$ used).

The flow time of the wind out to $r_{\rm inj}$ is about 120 years. The
ram pressure of the wind at the edge of the injection region is
$\rho_{\rm w}v_{\rm w}^{2}|_{\rm inj} = 1.8\times10^{-10}\,{\rm
  dyn\,cm^{-2}}$, which is more than $100\times$ higher than the
thermal pressure of the ambient medium. Nevertheless, we find that the
reverse shock initially tries to move back into the injection region,
and since the cell variables are overwritten at each step this causes
a small amount of mass and energy to be lost at early times (this loss
is avoided with the other wind launch methods). The reverse shock
eventually moves away from the injection region as the bubble grows
and becomes established.

\begin{figure}
\includegraphics[width=8.0cm]{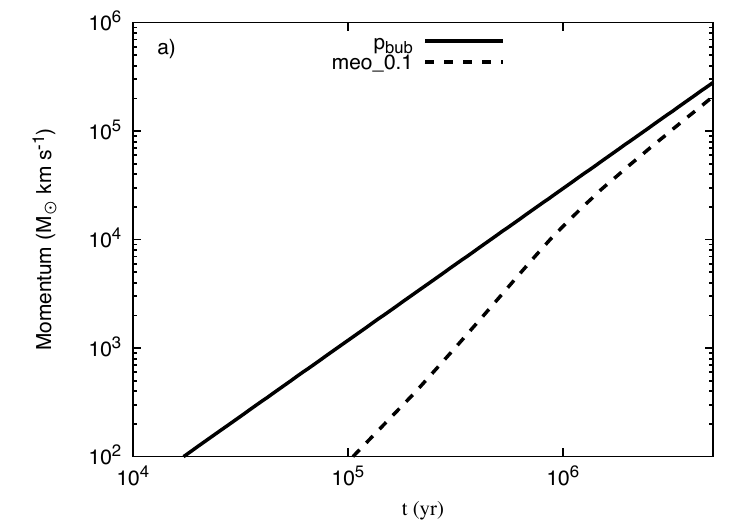}
\includegraphics[width=8.0cm]{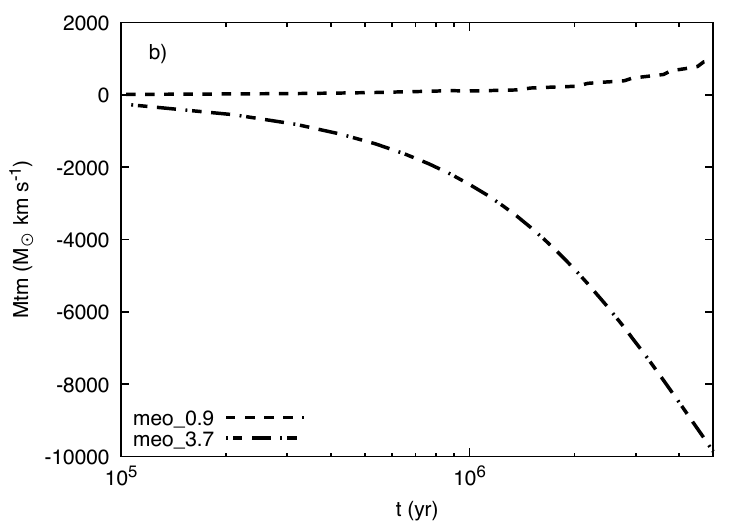}
\caption{The momentum of the bubble as a
  function of bubble age for simulations with various fixed resolutions. a) $dr =
0.025$\,pc (model {\it meo_0.1}; dashed line). Also shown is the
\citet{Weaver:1977} analytical prediction, which model {\em modx}
nearly approaches. b) $dr =
0.25$\,pc (model {\it meo_0.9}; dashed line) and $dr=1$\,pc (model {\it
  meo_3.7}; dot-dashed line).}
\label{fig:wbb_mtm2}
\end{figure}

Fig.~\ref{fig:wbb_mtm2}a) shows the bubble momentum from this
simulation. Compared to model {\it modx} the radial momentum is
significantly lower at early times as the bubble tries to establish
itself. In addition, the momentum of the bubble never fully catches up
to that in model {\it modx} or the analytical value, being still 25\%
lower after 5 Myrs.

\begin{figure}
\includegraphics[width=8.0cm]{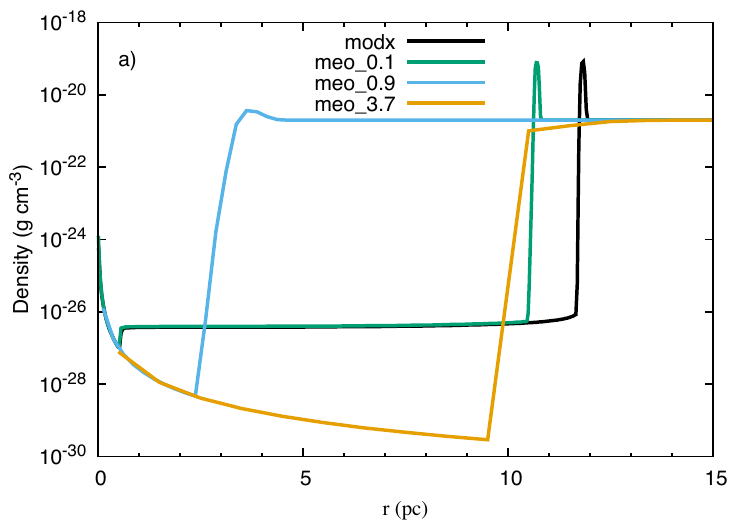}
\includegraphics[width=8.0cm]{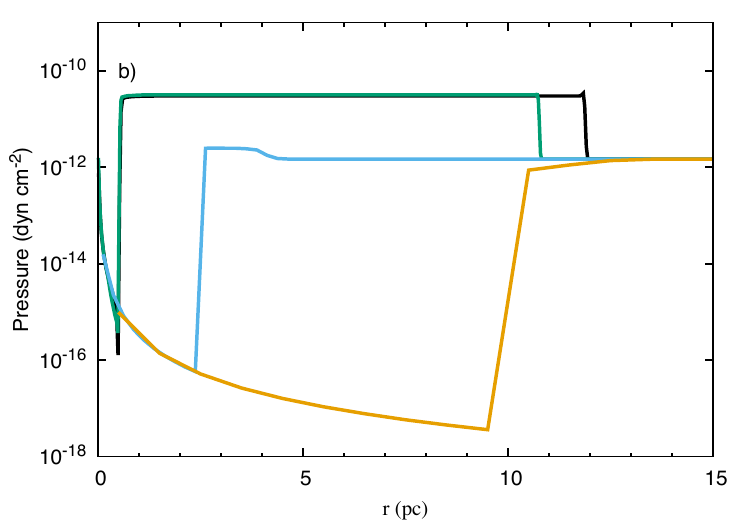}
\includegraphics[width=8.0cm]{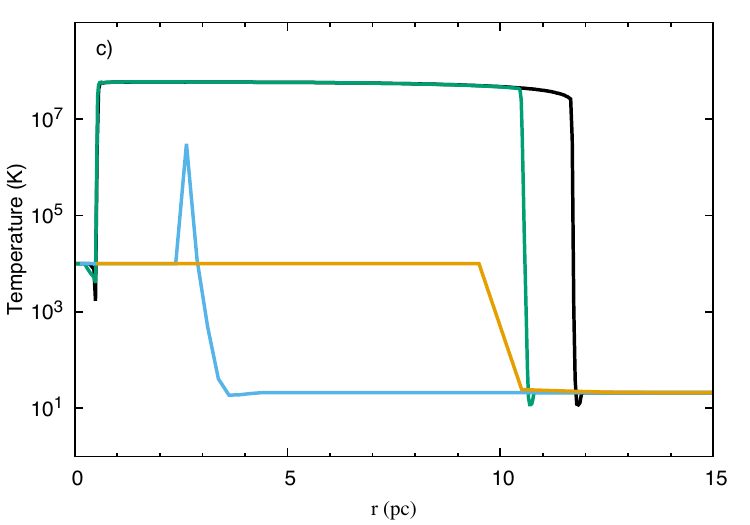}
\caption{Profiles of a) density; b) pressure; c) temperature at
  $t=5\,$Myr for models {\it modx}, {\it meo_0.1}, {\it meo_0.9} and
  {\it meo_3.7}. The ambient density, pressure and temperature values
  are visible on the far right of each plot. Models with $\chi = 0.1$,
  0.9 and 3.7 have $r_{\rm inj} = 0.25$\,pc, 2.5\,pc and 10\,pc,
  respectively. Note the differences in the positions of the reverse
  shock, contact discontinuity, and shell, and the amount of hot gas,
  as the resolution is varied.}
\label{fig:wbb_profile1}
\end{figure}

The situation becomes much worse if the grid cell size is further
increased. Fig.~\ref{fig:wbb_mtm2}b) shows the bubble momentum from
models {\it meo_0.9} with $dr=0.25$\,pc ($r_{\rm inj}=2.5$\,pc;
$\chi=0.93$) and {\it meo_3.7} with $dr=1.0$\,pc
($r_{\rm inj}=10$\,pc; $\chi=3.73$). The ram pressures of the winds at
the edge of the injection regions are now
$\rho_{\rm w}v_{\rm w}^{2}|_{\rm inj} = 1.8\times10^{-12}\,{\rm
  dyn\,cm^{-2}}$ and $1.0\times10^{-13}\,{\rm dyn\,cm^{-2}}$,
respectively. The former marginally exceeds $P_{\rm amb}$, which
allows for a small amount of hot gas to be created just outside of the
injection region (see Fig.~\ref{fig:wbb_profile1}). However, although
formally $r_{\rm inj} < r_{\rm inj,max}$, the shocked gas is not able
to do any useful $PdV$ work and the final radial momentum of
$\approx 10^{3}\,\Msol \kmps$ is only equal to the momentum injected
by the wind ($\beta \approx 1.0$). The bubble is not able to grow
outside of the injection region (and finished with a smaller radius
than in model {\it modx}), and nearly all of the surrounding medium
remains undisturbed.

For model {\it meo_3.7} with $dr=1.0$\,pc the circumstances are even
worse. Since $r_{\rm inj} > r_{\rm inj,max}$, the bubble is completely
quenched by the ambient pressure outside of the injection region. This
results in an inflow developing with negative radial momentum (see
Fig.~\ref{fig:wbb_mtm2}b), in complete disagreement with the higher
resolution models and analytical expectations.

Density, pressure and temperature profiles from the simulations at
$t=5$\,Myr are shown in Fig.~\ref{fig:wbb_profile1}. Model {\it modx}
is our high resolution reference, and has radii
$r_{\rm rs} \approx 0.53$\,pc, $r_{\rm cd} \approx 11.7$\,pc, and
$r_{\rm bub} \approx 11.9$\,pc. In model {\it meo_0.1}, the reverse
shock position and the shocked wind density, pressure and temperature
all agree with model {\it modx}. However, the shocked gas does not
extend as far out from the star, resulting in the contact
discontinuty, swept-up shell, and forward shock all appearing at too
small radii ($r_{\rm cd} \approx 10.5$\,pc and
$r_{\rm bub} \approx 10.8$\,pc).

In model {\it meo_0.9} the unshocked wind is forced to extend too far
from the star (past the position of the reverse shock in the reference
model). The smaller ram pressure at the edge of the injection region
is unable to grow a bubble and no significant hot gas is created (only
a small and narrow temperature spike at the edge of the injection
region - see Fig.~\ref{fig:wbb_profile1}b). In model {\it meo_3.7} the
unshocked wind is forced to extend out to a radius of 10\,pc, leading
to a wind density of $2.7\times10^{-30}\,{\rm g\,cm^{-3}}$ at the edge
of the injection region. This produces a ram pressure below
$P_{\rm amb}$, and backflow of gas.

\subsection{Other injection mechanisms}
In the previous subsection we examined the resolution dependence of
models with a momentum and energy overwrite (as in
Sec.~\ref{sec:meo}), which have a prefix of ``{\it meo}''.

We now examine models with energy injection (Sec.~\ref{sec:ei}), which
have a prefix of ``{\it ei}'', and models with momentum and energy
injection (Sec.~\ref{sec:mei}) which have a prefix of ``{\it
  mei}''. We explore the same 3 values of $\chi$ as before, with
models identified with the same ``{\it 0.1}'', ``{\it 0.9}'' or ``{\it
  3.7}'' postfix. We have also confirmed that when we use an expanding
grid, with $dr_{\rm 0}=10^{-5}$\,pc, different wind setups produce
identical bubbles with identical momenta (the momenta are within 0.1\%
at $t=5$\,Myr).

Note that we find that when an ``injection'' method (either {\it ei}
or {\it mei}) is used, in models with poor resolution (postfix {\it 0.9}
or {\it 3.7}), the nature of the resulting bubble is dependent on
whether additional constraints are placed on the courant number and
timestep. Bubbles are slightly more ``successful'' in these scenarios
if the initial courant number is set very low (e.g., $<10^{-4}$) and
slowly increased as the simulation progresses, and also if the global
timestep is in addition limited by the minimum net cooling time of the
gas in any cell. Models {\it ei_0.9}, {\it ei_3.7}, {\it mei_0.9} and {\it
  mei_3.7} all have these additional constraints.

\begin{table}
\begin{center}
  \caption[]{The models investigated. The columns show the model name,
    the wind launch method (see Sec.~\ref{sec:wind_launch_methods}),
    the radius of the injection region, the ratio of the injection
    region radius to its maximum possible value (Eq.~\ref{eq:chi}),
    and the momentum of the bubble and boost factor after
    5\,Myr. Model {\it modx} is the reference simulation that closely
    matches the analytical solution. All of the different wind launch
    methods result in exactly the same bubble properties in model {\it
      modx}, although in this paper we report on only the {\em meo}
    version. The maximum size that the injection region can be before
    any bubble is completely quenched is $r_{\rm inj,max} = 2.68$\,pc
    (Eq.~\ref{eq:rinj_max}), corresponding to $\chi=1.0$.}
\label{tab:models}
\begin{tabular}{llcccc}
\hline
  Model & Launch & $r_{\rm inj}$ & $\chi$ & $p_{\rm bub}$ & $\beta$\\
        & Method &   (pc)       &       & ($\Msol \,\kmps$) & \\
  \hline
  {\it modx} & meo & $10^{-4}$ & $3.73\times10^{-5}$ & $2.76\times10^{5}$ & 276\\
  {\it meo_0.1} & meo & 0.25 & 0.093 & $2.09\times10^{5}$ & 209 \\
  {\it meo_0.9} & meo & 2.5 & 0.93 & 966 & 0.97 \\ 
  {\it meo_3.7} & meo & 10 & 3.73 & $-9860$ & $-9.9$ \\ 
  {\it ei_0.1} & ei & 0.25 & 0.093 & $2.22\times10^{5}$ & 222 \\ 
  {\it ei_0.9} & ei & 2.5 & 0.93 & 523 & 0.52\\ 
  {\it ei_3.7} & ei & 10 & 3.73 & 151 & 0.15 \\ 
  {\it mei_0.1} & mei & 0.25 & 0.093 & $2.23\times10^{5}$ & 223 \\ 
  {\it mei_0.9} & mei & 2.5 & 0.93 & $5.30\times10^{4}$ & 53 \\ 
  {\it mei_3.7} & mei & 10 & 3.73 & 708 & 0.71 \\ 
\hline
\end{tabular}
\end{center}
\end{table}

\begin{figure}
\includegraphics[width=8.0cm]{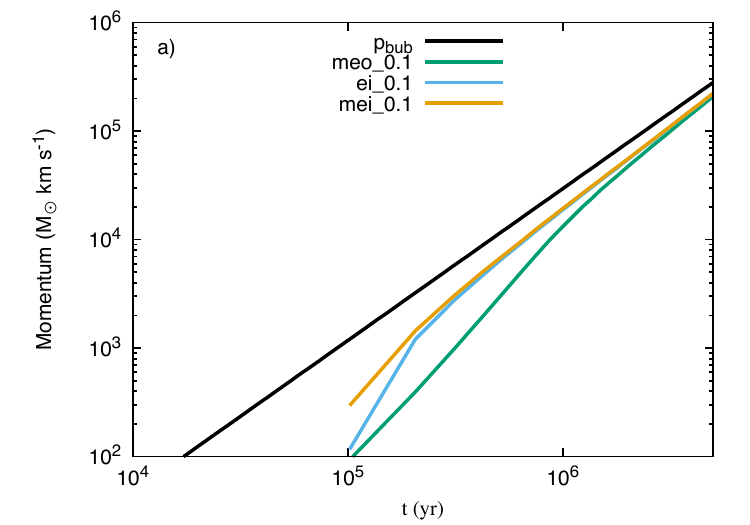}
\includegraphics[width=8.0cm]{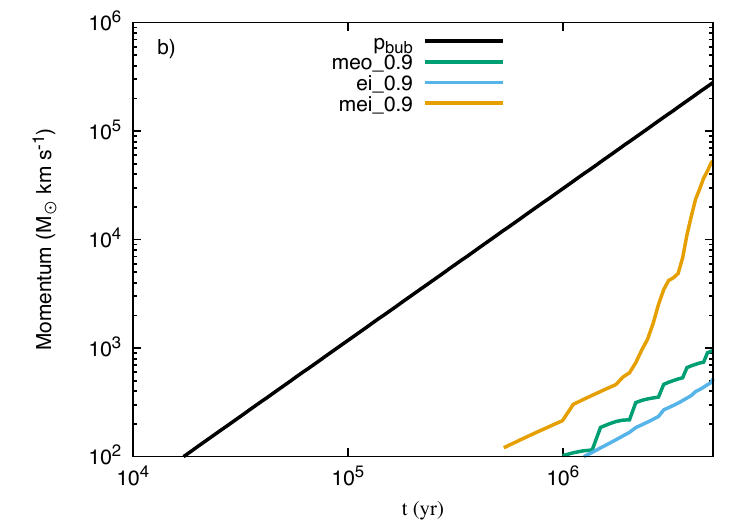}
\includegraphics[width=8.0cm]{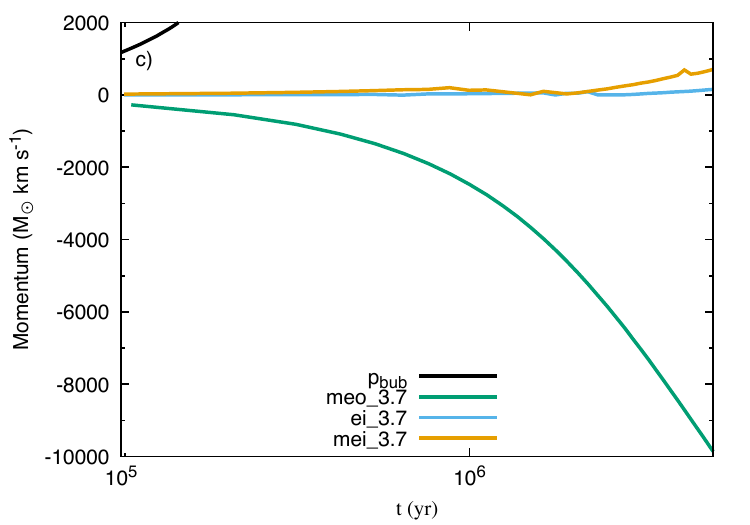}
\caption{The momentum of the bubble as a function of bubble age for
  simulations with a fixed grid and different wind injection
  mechanisms. a) $dr = 0.025$\,pc ($\chi=0.093$); b) $dr = 0.25$\,pc ($\chi=0.93$); c)
  $dr = 1$\,pc ($\chi=3.73$). Also shown is the \citet{Weaver:1977} analytical
  prediction.}
\label{fig:wbb_mtm_alp}
\end{figure}

\begin{figure}
\includegraphics[width=8.0cm]{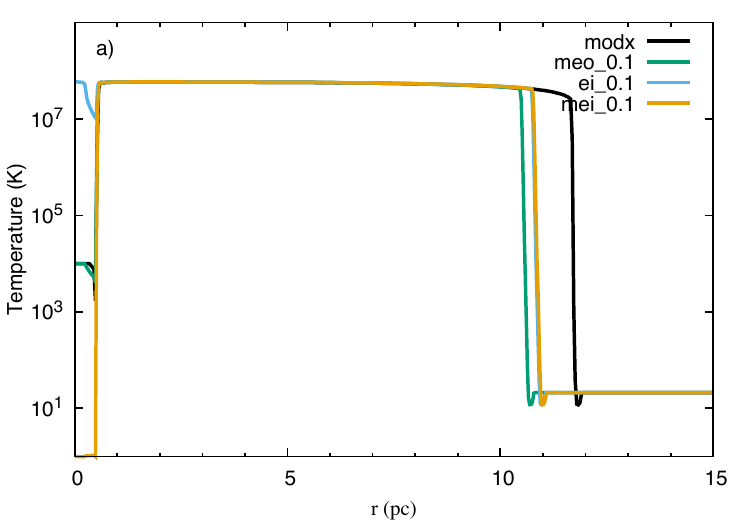}
\includegraphics[width=8.0cm]{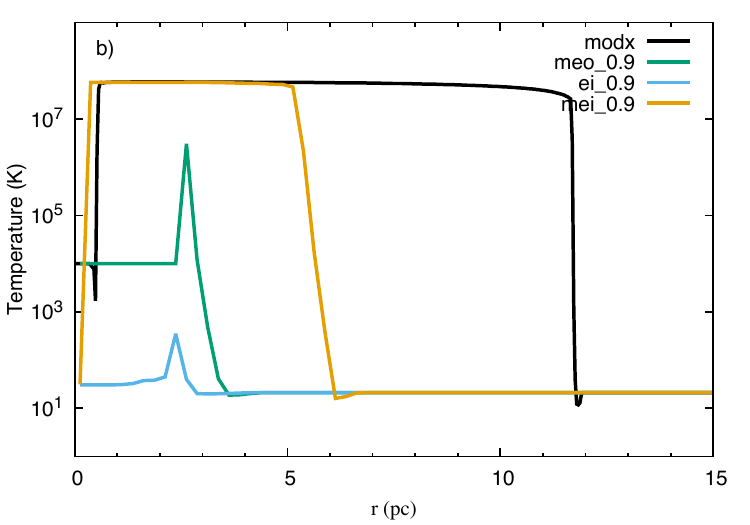}
\includegraphics[width=8.0cm]{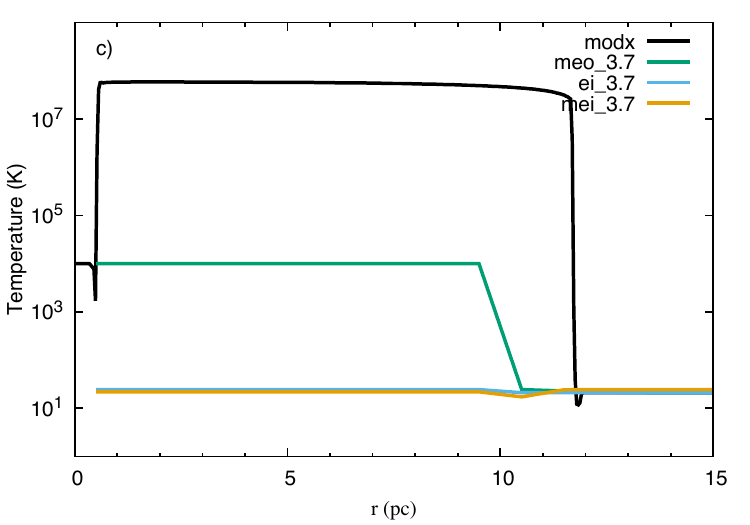}
\caption{Temperature profiles for simulations with different wind
  injection mechanisms and a fixed grid. a) $dr = 0.025$\,pc
  ($\chi=0.093$); b) $dr = 0.25$\,pc ($\chi=0.93$); c)
  $dr = 1$\,pc ($\chi=3.73$). Also shown is the temperature profile from model {\it
    modx}.}
\label{fig:wbb_temp_alp}
\end{figure}

Fig.~\ref{fig:wbb_mtm_alp}a) shows the bubble momentum from
simulations with a fixed $dr = 0.025$\,pc ($\chi=0.093$) and different
wind injection mechanisms. We see that the momentum rises most quickly
for method {\it mei}, then {\it ei}, and slowest for method {\it
  meo}. The final momentum produced for method {\it meo} is also about
6\% lower than obtained for methods {\it ei} and {\it mei}. This
behaviour is likely due to the reverse shock initially interacting
with the injection region in method {\it meo}, and results in a
slightly smaller bubble as shown in Fig.~\ref{fig:wbb_temp_alp}a)
which shows the corresponding temperature profiles. All 3 methods
produce final momenta which are 20\% lower (and bubbles that are
slightly smaller) than obtained for our reference model ({\it modx}).

Fig.~\ref{fig:wbb_mtm_alp}b) shows the bubble momentum from
simulations with a fixed $dr = 0.25$\,pc
($\chi=0.93$). Fig.~\ref{fig:wbb_temp_alp}b) shows the corresponding
temperature profiles. We see that methods {\it meo} and {\it ei} are
not able to generate a hot bubble. However, method {\it mei} is more
successful in this regard, and though the hot gas only extends out to
$\approx 5$\,pc (instead of the $\approx 12$\,pc seen in the reference
simulation {\it modx}), it is able to do significant $PdV$ work,
producing a final bubble momentum which is within a factor of 10 of
the analytical value (this is not the case if the additional
restrictions of an initially smaller courant number and timesteps
limited by the net cooling time are not implemented).

Fig.~\ref{fig:wbb_mtm_alp}c) shows the bubble momentum from
simulations with a fixed $dr = 1$\,pc
($\chi=3.73$). Fig.~\ref{fig:wbb_temp_alp}c) shows the corresponding
temperature profiles. No hot gas is generated using any of the
methods. With method {\it mei} the momentum of the gas
($700\,\Msol \kmps$) is less than the injected wind momentum
($10^{3}\,\Msol \kmps$), while method {\it ei} results in only
$150\,\Msol \kmps$ of momentum.

\begin{figure}
\includegraphics[width=8.0cm]{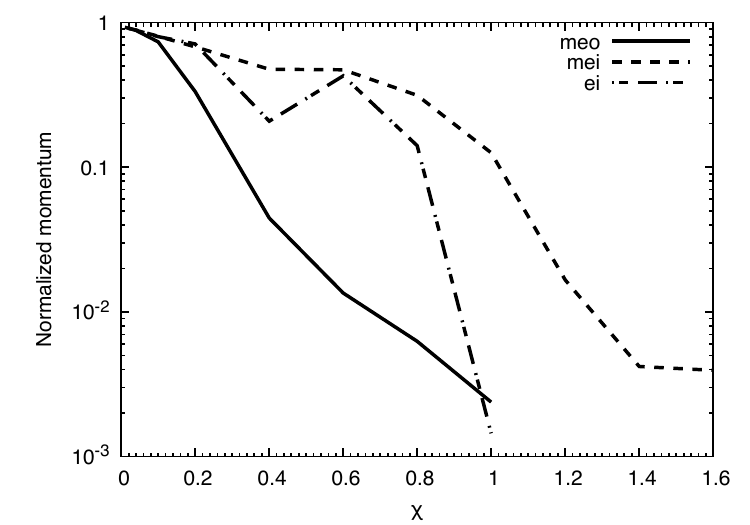}
\caption{The bubble momentum normalized to that from the
    reference model {\em modx},
  as a function of $\chi$, the ratio of the injection radius, $r_{\rm inj}$, to
  the maximum injection radius, $r_{\rm inj,max}$. Results for the
  three different wind launch methods are shown. The momenta are all
  measured at a bubble age of 5\,Myr.}
\label{fig:wbb_mtmvschi}
\end{figure}

In Table~\ref{tab:normalizedmtm} and Fig.~\ref{fig:wbb_mtmvschi} we
show the bubble momentum, normalized to that from {\em modx}, as a
function of the ratio $\chi = r_{\rm inj}/r_{\rm inj,max}$. We see
that all 3 methods capture $75-80$\% of the expected bubble momentum
when $\chi =0.1$. This value of $\chi$ marks a turning point for the bubble
momentum when using method {\it meo}, which displays a power-law
decline between $\chi=0.2-0.8$. Less than 1 per cent of the expected
bubble momentum is attained when $\chi = 0.8$. In contrast, with
methods {\it mei} and {\it ei} the bubble momentum remains closer to
the reference model when $\chi > 0.1$, although a sharp decline eventually
occurs. Method {\it mei} is still able to create a hot bubble when
$\chi=1.0$, though this ability disappears for values of
$\chi \gtsimm 1.0$.

Fig.~\ref{fig:wbb_mtmvschi} shows that we must have $\chi \ltsimm 0.1$
in order to obtain a bubble momentum within $20-25$\% of the
analytical (or reference model) value. To be within 10\% of the
analytical value requires $\chi \ltsimm 0.02$.

\begin{table}
\begin{center}
\caption[]{The radial momentum of the bubble, as a function of $\chi$,
  for the 3 wind launch models investigated. The radial momentum has
  been normalized by the value from the high resolution reference model ({\it
    modx}). The momentum is measured at $t=5\,$Myr.}
\label{tab:normalizedmtm}
\begin{tabular}{llll}
\hline
  $\chi$ & {\em meo} & {\em ei} & {\em mei} \\
  \hline
  0.01 & 0.928 & 0.928 & 0.928 \\
  0.02 & 0.913 & 0.913 & 0.914 \\
  0.04 & 0.882 & 0.886 & 0.888 \\
  0.10 & 0.736 & 0.799 & 0.804\\
  0.20 & 0.334 & 0.711 & 0.680 \\
  0.40 & 0.0446 & 0.208 & 0.475 \\
  0.60 & 0.0135 & 0.428 & 0.472 \\
  0.80 & 0.0063 & 0.141 & 0.314\\
  1.00 & 0.0028 & 0.0015 & 0.126 \\
\hline
\end{tabular}
\end{center}
\end{table}

\section{Discussion}
\label{sec:discuss}
In Sec.~\ref{sec:results} we find that the momentum of the bubble can
be significantly underestimated in simulations where the numerical
resolution is insufficient. For a bubble to be created we
require $\chi = r_{\rm inj}/r_{\rm inj,max} < 1$. However, for a
bubble to match the analytical solution reasonably well requires
$\chi < 0.1$, and still higher resolutions are required to obtain a
momentum boost within 10 per cent of the analytical solution.

In the literature, simulations with a variety of values for $\chi$ can
be found. In their stellar feedback paper, \citet{Rogers:2013} used a
resolution $dx=0.0625$\,pc, and an injection region radius
$r_{\rm inj}=6\,dx=0.375$\,pc. For the first 4\,Myr of the simulation,
the wind momentum injection rate was
$\dot{p}_{\rm wind}=1.14\times10^{28}\,{\rm g\,cm\,s^{-2}}$. The
average ambient pressure within the GMC clump that the stellar
feedback blows into was $2.8\times10^{-13}\,{\rm dyn\,cm^{-2}}$. This
gives $r_{\rm inj,max}=18.5$\,pc and $\chi \approx 0.02$. Therefore,
the bubble that forms is initially highly overpressured and has no
problem in growing. We expect that nearly all of the initial growth in
the bubble momentum will have been captured (although at later times
the bubble expands off the grid).

An example where the wind injection has not been sufficiently resolved
is \citet{Geen:2015}. In this paper feedback from a single 15\,$\Msol$
star into a variety of ambient densities and temperatures, spanning
the range ($n=0.1\,{\rm cm^{-3}},T=62\,$K) to
($n=100\,{\rm cm^{-3}},T=8.2\,$K), is considered. The wind and the
ionizing radiation from the star are both considered and it is
concluded that the stellar wind has negligible impact. We estimate the
wind parameters as $\Mdot \approx 10^{-8}\,\Msolpyr$ and
$v_{\rm w}\approx 1000\,\kmps$ (for solar abundances). For their densest ambient medium we
estimate $r_{\rm inj,max}\approx3$\,pc, while for their lowest density
medium we estimate $r_{\rm inj,max}\approx35$\,pc. As they use
$r_{\rm inj}=12$\,pc, they either fail to launch a bubble at all
($\chi \approx 4$), or the bubble expansion is severely compromised
($\chi \approx 0.34$). Therefore, their claim that the stellar wind has
negligible impact (see their Fig.~11) should be revisited.

\citet{Haid:2018} study feedback into initially warm and ionized gas
(WIM; $\rho = 2.1\times10^{-25}\,{\rm g\,cm^{-3}}$ and
$T = 10^{4}\,$K) and into cold, predominantly neutral gas (CNM;
$\rho = 2.1\times10^{-22}\,{\rm g\,cm^{-3}}$ and $T = 20\,$K). Both
their WIM and CNM have $P/k = 10^{3}\,{\rm K \,cm^{-3}}$ and an
injection radius of 2.4\,pc is used. We estimate that
$r_{\rm inj,max}=1.8$, 13 and 52\,pc for their models with stars of
mass $M_{*}=12$, 23 and $60\,\Msol$, respectively, giving
$\chi \approx 1.3$, 0.18 and 0.05. Thus we expect the bubble around
their $12\,\Msol$ star to be completely missing, and it should be
significantly compromised around their $23\,\Msol$ star. We believe
that only their $60\,\Msol$ star models blow a bubble that would
closely match higher resolution models. Our expectations appear to be
valid: their Fig.~3 shows no evidence of hot gas in their $12\,\Msol$
CNM model, while their Figs.~3 and~5 reveal the presence of a reverse
shock only for the 23 and 60\,$\Msol$ models in the WIM, and for the
60\,$\Msol$ model in the CNM. Thus, the authors claims should also be
reexamined.

Our own work is also not immune from these issues. Although the
simulations with the $40-120\,\Msol$ stars in
\citet{Wareing:2017a,Wareing:2017b} have $\chi \ltsimm 0.1$ and
vigorously inflate bubbles, the wind injection in the $15\,\Msol$ star
simulations is estimated to have $\chi \approx 0.5-0.6$. Although hot
($\sim 10^{8}\,$K) shocked stellar wind gas is present, and flows some
distance from the injection region, a strong reverse shock is not
always visible. Thus, the stellar wind impact is likely to be strongly
underestimated in the $15\,\Msol$ star simulations.

We would like to stress that the papers discussed in this section are
simply ones that we are familiar with in the literature - there are
likely to be other papers with similar issues. It also remains the
case that even if the feedback from lower mass (e.g., $12-15\,\Msol$)
stars has not always been modelled with sufficient resolution, the
winds from such stars may still be too weak to strongly affect their
environment. If this is so, the conclusions from these papers will
still stand.

Our results are presented for a uniform medium and show that the ram
and/or thermal pressure at the edge of the injection region must
significantly exceed the ambient pressure in order to correctly
inflate the bubble. In reality, bubbles usually interact with a highly
inhomogeneous medium. Previous work in the literature
\citep*[e.g.,][]{Rogers:2013,Kim:2017,Lancaster:2021b} shows that the
size and structure of the bubble depends on the number, size, density
contrast and distribution of the clouds. The location of the reverse
shock may in places be determined by the position of individual
clouds. However, since $r_{\rm inj,max}$ depends on $P_{\rm amb}$ (not
$\rho_{\rm amb}$), the requirement that $\chi$ needs to be signicantly
less than unity likely remains valid.

We note two further points. While we have only focussed on the thermal
pressure of the surroundings, the position of the reverse shock will
in fact depend on the {\em total} pressure (thermal + magnetic +
turbulent + cosmic ray). When there is significant non-thermal
pressure, the total ambient pressure should be used when evaluating
$r_{\rm inj,max}$. Finally, if one is only concerned with
momentum-driven feedback no constraint exists on the value of
$\chi$. This is because no extra momentum is being created through
$PdV$ work by a hot bubble.  In such cases it should be possible to
use very low numerical resolution (though this has other consequences,
such as the ability to resolve structures and flows at a particular
scale).

\section{Summary and conclusions}
\label{sec:summary}
We have examined the numerical resolution requirements to blow
energy-driven stellar wind bubbles in a uniform medium. We have
determined a maximum radius for the wind injection region,
$r_{\rm inj,max}$, above which a bubble will not usually grow. This
applies to all 3 wind injection mechanisms studied.  If
$\chi = r_{\rm inj}/r_{\rm inj,max} < 1$ is only marginally satisfied,
the resulting bubble will be only marginally overpressured and unable
to generate the large momentum boost that it should.

In order for the bubble momentum to match analytical predictions, the
very early growth of the bubble must be captured as accurately as
possible which requires very high resolution.  To ensure this, the
flow time of the wind out to the edge of the injection region should
be significantly less than the time at which the free-flowing wind and
bubble momenta are equal ($t_{\rm eq}$). This requires that
$r_{\rm inj} \ll t_{\rm eq}/v_{\rm w}$. If $r_{\rm inj}$ is
appropriately chosen, the two-shocks that initially develop when using
method {\it meo} should both move outwards. This ensures that no mass,
momentum, or energy is lost from the simulation. All 3 injection
methods yield the same bubble properties and momentum for such small
values of $r_{\rm inj}$.

As $\chi$ is increased, the bubble loses more and more momentum, due
to the absence of the high initial pressures that actual bubbles have.
When $0.1 < \chi \leq 1.0$ the momentum and energy ({\it mei}) wind
injection method outperforms the other methods (restrictions on the
courant number and radiative cooling limits on the timestep
aside). However, if $\chi = 0.1$, we find that $20-25$\% of the bubble
momentum is still missed. To be within 10\% of the momentum from the
reference model requires $\chi \ltsimm 0.02$, in which case all wind
injection methods perform similarly without the need for such
additional restrictions.

This paper highlights that the injection region of the stellar
wind must be adequately resolved. Because our calculations are one
dimensional, restrict cooling at unresolved interfaces, and do not
include thermal conduction or explicit mixing of hot and cold phases,
the cooling of the hot gas inside the bubble is minimised (and the
momentum of the bubble is maximised). The actual impact of these
restrictions and processes is still to be determined.

\section*{Acknowledgements}
We thank the referee for their helpful comments. JMP was supported by
grant ST/P00041X/1 (STFC, UK).

\section*{Data Availability}
The data underlying this article are available in the Research
Data Leeds Repository, at \url{https://doi.org/10.5518/1046}.

%%%%%%%%%%%%%%%%%%%%%%%%%%%%%%%%%%%%%%%%%%%%%%%%%%



%\bibliographystyle{mnras}

\begin{thebibliography}{99}
\bibitem[\protect\citeauthoryear{Chevalier \& Clegg}{1985}]{Chevalier:1985}
Chevalier R.~A., Clegg A.~W., 1985, Nature, 317, 44
\bibitem[\protect\citeauthoryear{Chevance et al.}{2020}]{Chevance:2020}
Chevance M., et al., 2020, MNRAS, 493, 2872 % more than 20 authors  
\bibitem[\protect\citeauthoryear{Chevance et al.}{2021}]{Chevance:2021}
Chevance M., et al., 2021, MNRAS, submitted (arXiv::2010.13788) % more than 20 authors  
\bibitem[\protect\citeauthoryear{Dale et al.}{2014}]{Dale:2014}
Dale J.~E., Ngoumou J., Ercolano B., Bonnell I.~A., 2014, MNRAS, 442, 694
\bibitem[\protect\citeauthoryear{Dinnbier \& Walch}{2020}]{Dinnbier:2020}
Dinnbier F., Walch S., 2020, MNRAS, 499, 748
\bibitem[\protect\citeauthoryear{Dyson \& Williams}{1980}]{Dyson:1980}  
Dyson J.~E., Williams D.~A., 1980, The Physics of the Interstellar
Medium. Halsted Press, New York
\bibitem[\protect\citeauthoryear{El-Badry et al.}{2019}]{el-badry:2019}
El-Badry K., Ostriker E.~C., Kim C.-G., Quataert E., Weisz D.~R.,
2019, MNRAS, 490, 1961  
\bibitem[\protect\citeauthoryear{Garc\'{i}a-Segura \& Franco}{1996}]{Garcia:1996}
Garc\'{i}a-Segura G., Franco J., 1996, ApJ, 469, 171  
\bibitem[\protect\citeauthoryear{Gatto et al.}{2017}]{Gatto:2017}
Gatto A., et al., 2017, MNRAS, 466, 1903  
\bibitem[\protect\citeauthoryear{Geen et al.}{2015}]{Geen:2015}
Geen S., Rosdahl J., Blaizot J., Devriendt J., Slyz A., 2015, MNRAS,
448, 3248
\bibitem[\protect\citeauthoryear{Geen et al.}{2021}]{Geen:2021}
Geen S., Bieri R., Rosdahl J., de Koter A., 2021, MNRAS (arXiv:2009.08742)
\bibitem[\protect\citeauthoryear{Girichidis et al.}{2018}]{Girichidis:2018}
Girichidis P., Seifried D., Naab T., Peters T., Walch S., W\"{u}nsch
R., Glover S.~C.~O., Klessen R.~S., 2018, MNRAS, 480, 3511 
\bibitem[\protect\citeauthoryear{Gnat \& Ferland}{2012}]{Gnat:2012}
Gnat O., Ferland G.~J., 2012, ApJS, 199, 20
\bibitem[\protect\citeauthoryear{Grevesse et al.}{2010}]{Grevesse:2010}
Grevesse N., Asplund M., Sauval A.~J., Scott P., 2010, Ap\&SS, 328, 179  
\bibitem[\protect\citeauthoryear{Haid et al.}{2018}]{Haid:2018}
Haid S., Walch S., Seifried D., W\"{u}nsch R., Dinnbier F., Naab T., 2018,
MNRAS, 478, 4799
\bibitem[\protect\citeauthoryear{Kaastra}{1992}]{Kaastra:1992}
Kaastra J.~S., 1992, An X-ray Spectral Code for Optically Thin Plasmas.
Internal SRON-Leiden Report
\bibitem[\protect\citeauthoryear{Kim, Ostriker \& Raileanu}{Kim et al.}{2017}]{Kim:2017}
Kim C.-G., Ostriker E.~C., Raileanu R., 2017, ApJ, 834, 25
\bibitem[\protect\citeauthoryear{Koyama \& Inutsuka}{2000}]{Koyama:2000}
Koyama H., Inutsuka S.-I., 2000, ApJ, 532, 980
\bibitem[\protect\citeauthoryear{Kruijssen et al.}{2019}]{Kruijssen:2019}
Kruijssen J.~M.~D., et al., 2019, Nature, 569, 519 %10 authors
\bibitem[\protect\citeauthoryear{Krumholz, Stone \& Gardiner}{Krumholz
    et al.}{2007}]{Krumholz:2007}
Krumholz M.~R., Stone J.~M., Gardiner T.~A., 2007, ApJ, 671, 518
\bibitem[\protect\citeauthoryear{Lancaster et al.}{2021a}]{Lancaster:2021a}
Lancaster L., Ostriker E.~C., Kim J.-G., Kim C.-G., 2021, ApJ, 914, 89
\bibitem[\protect\citeauthoryear{Lancaster et al.}{2021b}]{Lancaster:2021b}
Lancaster L., Ostriker E.~C., Kim J.-G., Kim C.-G., 2021, ApJ, 914, 90
\bibitem[\protect\citeauthoryear{Mewe, Kaastra \& Liedahl}{Mewe et al.}{1995}]{Mewe:1995}
Mewe R., Kaastra J.~S., Liedahl D.~A., 1995, Legacy, 6, 16
\bibitem[\protect\citeauthoryear{Peters et al.}{2017}]{Peters:2017}
Peters T., et al., 2017, MNRAS, 466, 3293
\bibitem[\protect\citeauthoryear{Pittard}{2013}]{Pittard:2013}
Pittard J.~M., 2013, MNRAS, 435, 3600
\bibitem[\protect\citeauthoryear{Rathjen et al.}{2021}]{Rathjen:2021}  
Rathjen T.-E., Naab T., Girichidis P., Walch S., W\"{u}nsch R.,
Dinnbier F., Seifried D., Klessen R.~S., Glover S.~C.~O., 2021, MNRAS,
504, 1039
\bibitem[\protect\citeauthoryear{Rogers \& Pittard}{2013}]{Rogers:2013}
Rogers H., Pittard J.~M., 2013, MNRAS, 431, 1337
\bibitem[\protect\citeauthoryear{Shetty \& Ostriker}{2012}]{Shetty:2012}
Shetty R., Ostriker E.~C., 2012, ApJ, 754, 2
\bibitem[\protect\citeauthoryear{Smith, Sijacki \& Shen}{Smith et al.}{2019}]{Smith:2019}
Smith M.~C., Sijacki D., Shen S., 2019, MNRAS, 485, 3317
\bibitem[\protect\citeauthoryear{Sutherland}{2010}]{Sutherland:2010}
Sutherland R.~S., 2010, Ap\&SS, 327, 173
\bibitem[\protect\citeauthoryear{Wareing et al.}{2016}]{Wareing:2016}
Wareing C.~J., Pittard J.~M., Falle S.~A.~E.~G., Van Loo S., 2016,
MNRAS, 459, 1803
\bibitem[\protect\citeauthoryear{Wareing, Pittard \& Falle}{Wareing et al.}{2017a}]{Wareing:2017a}
Wareing C.~J., Pittard J.~M., Falle S.~A.~E.~G., 2017a, MNRAS, 465, 2757
\bibitem[\protect\citeauthoryear{Wareing, Pittard \& Falle}{Wareing et al.}{2017b}]{Wareing:2017b}
Wareing C.~J., Pittard J.~M., Falle S.~A.~E.~G., 2017b, MNRAS, 470, 2283
\bibitem[\protect\citeauthoryear{Wareing et al.}{2018}]{Wareing:2018}
Wareing C.~J., Pittard J.~M., Wright N.~J., Falle S.~A.~E.~G., 2018, MNRAS, 475, 3598
\bibitem[\protect\citeauthoryear{Weaver et al.}{1977}]{Weaver:1977}
Weaver R., McCray R., Castor J., Shapiro P., Moore R., 1977, ApJ, 218, 377
\bibitem[\protect\citeauthoryear{Wunsch et al.}{2008}]{Wunsch:2008}
W\"{u}nsch R., Tenorio-Tagle G., Palou\u{s} J., Silich S., 2008, ApJ,
683, 683  
\end{thebibliography}


%%%%%%%%%%%%%%%%%%%%%%%%%%%%%%%%%%%%%%%%%%%%%%%%%%

%%%%%%%%%%%%%%%%% APPENDICES %%%%%%%%%%%%%%%%%%%%%

%\appendix

% Don't change these lines
\bsp	% typesetting comment
\label{lastpage}
\end{document}